\documentclass[twocolumn,twocolappendix]{aastex631}

\usepackage{multirow}
\usepackage{makecell}

\submitjournal{PASP}

\shorttitle{Observation Strategy for Accurate Dark-Sky Flats}
\shortauthors{Byun et al.}
\graphicspath{{./}{figures/}}

\begin{document}

\title{K-DRIFT Preparation: Experimental Verification of an Observation Strategy for Accurate Dark-Sky Flats}

\correspondingauthor{Woowon Byun}
\email{wbyun87@gmail.com}

\author[0000-0002-7762-7712]{Woowon Byun}
\affiliation{Korea Astronomy and Space Science Institute, Daejeon 34055, Republic of Korea}

\author[0000-0001-9561-8134]{Kwang-Il Seon}
\affiliation{Korea Astronomy and Space Science Institute, Daejeon 34055, Republic of Korea}
\affiliation{Department of Astronomy and Space Science, University of Science and Technology, Daejeon 34113, Republic of Korea}

\author[0000-0002-9434-5936]{Jongwan Ko}
\affiliation{Korea Astronomy and Space Science Institute, Daejeon 34055, Republic of Korea}
\affiliation{Department of Astronomy and Space Science, University of Science and Technology, Daejeon 34113, Republic of Korea}



\begin{abstract}

Despite its scientific importance, the low-surface-brightness universe has yet to be fully explored due to various systematic uncertainties that affect the achievable surface-brightness limit. Reducing these uncertainties requires very accurate data processing. The dark-sky flat is a widely used calibration frame for accurate flat-field correction, generated by combining the sky background from science images. However, the night sky will likely contain complex local fluctuations, thus may still lead to photometric errors in data calibrated with dark-sky flats. To address this concern, we conduct mock observations with semi-realistic sky simulation data and evaluate observation strategies to mitigate the impact of the fluctuating sky background. Our experiments consider two representative sky conditions (clear and dirty) and perform intensive comparative analysis on two observation methods (offset and rolling). Our findings suggest that the rolling dithering method, which incorporates the operation of camera rotation into conventional dithering, can provide more accurate dark-sky flats. Finally, we discuss the broader implications of this method through additional experiments examining several factors that may affect the imaging quality of observational data. 

\end{abstract}

\keywords{Astronomical techniques (1684); Astronomy data reduction (1861); Sky brightness (1462); Wide-field telescopes (1800)}


\section{Introduction} \label{sec:intro}
Galaxies and galaxy groups/clusters have evolved through interactions and mergers \cite[][]{1978MNRAS.183..341W,1991ApJ...379...52W}, so circum- and inter-galactic spaces are expected to be full of very faint structures, i.e., tidally disrupted remnants \cite[][]{1988ApJ...331..699B,1992ApJ...399L.117H,2007ApJ...666...20P,2008ApJ...684.1062F,2008ApJ...689..936J,2013MNRAS.434.3348C,2014MNRAS.444..237P}. Because of their long dynamical lifespan, these remnants can provide hints at the mass assembly histories of their host galaxies and themselves \cite[see][]{2001ApJ...548...33B,2008MNRAS.391...14D,2008ApJ...689..936J}. Typically, tidal remnants have extremely low surface brightness (LSB), thus detecting them requires reaching a surface-brightness limit of 27 mag arcsec$^{-2}$ or fainter \cite[see][]{2011ApJ...739...20C,2012arXiv1204.3082B}. This is why we have not yet fully explored the LSB universe. 

Since LSB observational studies aim to detect light fainter than 1\% of the night sky brightness---which corresponds to the typical surface-brightness level of dwarf galaxies, tidal remnants, and stellar halos---collecting as much light as possible is essential. However, an even more important challenge is controlling the uncertainties that may impact the detection and analysis of LSB features. These uncertainties generally involve systematic errors, primarily arising from stray light and data processing. For this reason, telescopes with small but simple optics are more suitable for LSB observations than large telescopes with complex optical designs. Notable examples include the Dragonfly Telephoto Array \cite[][]{2014PASP..126...55A}, which utilizes multiple small-aperture lenses, and the Stellar Tidal Streams Survey \cite[][]{2019hsax.conf..146M}, which employs a network of modest-sized amateur telescopes. 

The Korea Astronomy and Space Science Institute (KASI) has been developing novel telescopes optimized for LSB observations. The KASI Deep Rolling Imaging Fast Telescope (K-DRIFT; J. Ko et al., in preparation) is a wide-field optical telescope for exploring the southern hemisphere sky in the wavelength range from  3000 to 7000 {\AA}. This telescope minimizes stray light to negligible levels by employing an off-axis freeform three-mirror system without any obscuring structures in the optical path \cite[][]{10.1117/12.2023433,2020PASP..132d4504P}. Consequently, achieving deep photometric depth primarily depends on reducing errors in data processing. Of these processes, accurate flat-field correction---which compensates for pixel-to-pixel sensitivity variations in imaging sensors such as a charge-coupled device (CCD) or a complementary metal-oxide-semiconductor (CMOS) sensor---is a critical step for LSB research. 

Generally, three types of master flats can be used for flat-field correction: (1) a twilight flat, (2) a dome flat, and (3) a dark-sky flat.\footnote{This term is often used interchangeably with the term super-sky flat. A dark-sky flat is more appropriate for relatively short, hour-scale observations, while a super-sky flat is suitable for extensive data obtained over multiple nights.} A twilight flat is one of the most commonly used calibration frames, observing the twilight sky that is believed to be evenly illuminated. It can be captured quickly due to the relatively bright sky, but securing enough frames would be tricky within a limited time if multiple filters are used. A dome flat is more convenient because it can be obtained anytime, regardless of weather conditions, and also offers a stable and controllable illumination source. However, implementing a highly uniform illumination source for specific telescopes requires careful instrumental design, often including auxiliary devices \cite[see][]{2004AJ....127.3642Z,2013PASP..125.1277M}. 

Dark-sky flats are generated by combining the sky background of science images after a dedicated masking process, which requires lots of time and effort. Because this frame retains the conditions of the target observation, flat-field correction using it can be made more accurate. For example, it can eliminate camera artifacts that vary depending on the exposure time. It is also possible to remove fringe patterns that commonly occur from long wavelength observations without employing a separate fringe frame \cite[see][]{2012PASP..124..263H}. 

However, dark-sky flats are still considered imperfect because of the non-uniformity of the night sky background. For instance, \cite{1996PASP..108..944C} reported that sky gradients as large as 1\% typically occur within a degree-scale field of view (FoV), and \cite{2014SPIE.9149E..2HW} also found similar gradient levels even in regions which are theoretically expected to have minimal gradient. This highlights the necessity of mitigating such non-uniformity, particularly for deep and wide-field imaging surveys. Additionally, although beyond the scope of this paper, slight angular displacements between instruments and light sources can introduce further non-uniform illumination patterns \cite[][]{1997PASP..109.1269W}, and irregularities in the sensor's pixel grid may also impact flat-field correction \cite[][]{2017PASP..129h4502B}.

One known method for obtaining an accurate master flat is to use drift-scan mode. This method records the average sky brightness along a linear trail as the sky background travels across the detector in sub-pixel units, which is believed to reflect the detector's non-uniform sensitivity well. However, while the sampling effect of each pixel through the scan can be robust, it cannot completely eliminate the impact of variations in sky brightness, particularly global gradients. 

For this reason, more optimized observation strategies are required to actively mitigate the effects of sky brightness variation. By the way, for wide-field observations in drift-scan mode, curvature correction is often achieved by rotating the detector along the great circle. Inspired by this, K-DRIFT employs the capability of a rotating detector and leverages it to develop an observation strategy to improve the accuracy of dark-sky flats. To validate the effectiveness of K-DRIFT's approach, we examined the accuracy of dark-sky flats under various observational conditions and methods. 

This paper outlines several experiments on observation strategies aimed at obtaining the most accurate dark-sky flat. Section \ref{sec:method} describes the details of night sky simulations and mock observations conducted with them. Section \ref{sec:result} compares the accuracy of dark-sky flats and evaluates which approach best reduces flat-fielding error. Section \ref{sec:disc} discusses various factors that may affect the accuracy of dark-sky flats, and Section \ref{sec:sum} summarizes the overall findings from the experiments. 

\section{Methods} \label{sec:method}
Accurate data processing is crucial for achieving a surface-brightness limit as faint as 30 mag arcsec$^{-2}$, which we aim to reach. Since dark-sky flats are generated by combining the sky background pixels in science images, their accuracy depends on the condition of the night sky at the time of observation. Therefore, it is necessary to develop an observation strategy for obtaining dark-sky flats that minimize the impact of varying sky backgrounds while avoiding significant time loss. 

Note that dark-sky flats generated under the darkest, clearest, and most stable night sky conditions can alternatively be used as long-term calibration frames. However, since the detector's performance is likely to change over time and night sky conditions cannot always be the same, this study premises that data obtained on a given day are processed using a dark-sky flat generated from data on the same day. 

To obtain sky frames under various conditions, we simulated a semi-realistic night sky, incorporating only the prominent light sources to simplify the model. Using this simulation, we performed mock observations and evaluated the accuracy of the resulting dark-sky flats under different observation strategies. Further details are described in the following sections. 

\subsection{Night Sky Simulation} \label{sec:sim}
\subsubsection{Sky Background} \label{sec:sim:bkg}
The night sky contains various light sources, which introduce spatial and temporal variations in brightness. While moonlight and light pollution can sometimes be significant, they are excluded from this study as their effects can be avoided by adjusting the observation position and schedule. Three other significant components generally govern the brightness of the night sky: airglow (AG), zodiacal light (ZL), and diffuse Galactic light (DGL). 

\begin{figure*}[t!]
\includegraphics[width=\linewidth]{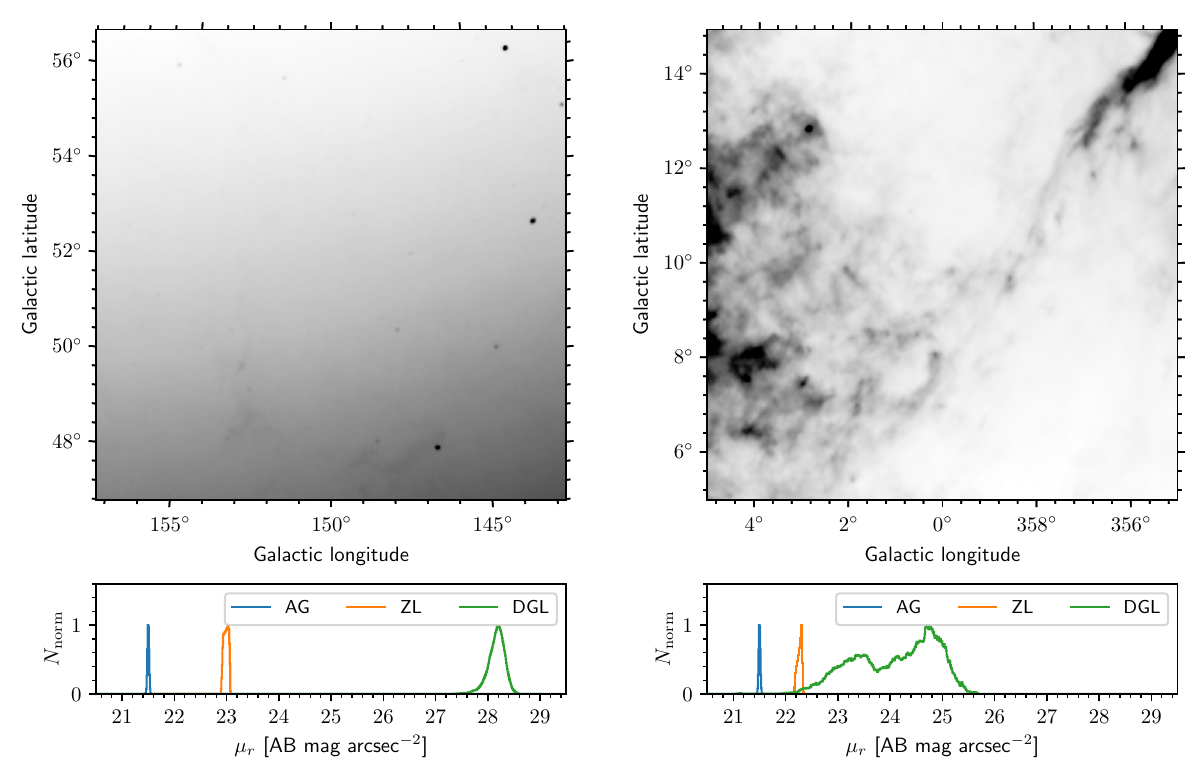}
\caption{Top: the 2D maps of two contrasting sky backgrounds; the clear (left) and dirty (right) skies. Bottom: the surface-brightness distributions of the three light sources for each sky. The SB of AG is presumed to be $\mu_r\sim21.5\pm0.05$ mag arcsec$^{-2}$. \label{fig:f1}}
\end{figure*}

AG is the brightest component in the sky background, resulting from a combination of emissions at various wavelengths across different layers of Earth's atmosphere. AG exhibits considerable fluctuations on both short and long timescales, driven by changes in the atmosphere and solar activity \cite[][]{1998A&AS..127....1L,2012A&A...543A..92N}. The dynamic range of long-term changes, including night-time and seasonal variations, can reach tens of percent, depending on the wavelength \cite[][]{2007PASP..119..687K}. AG also changes with the zenith angle, which is why the K-DRIFT survey will primarily be conducted near the zenith to minimize the spatial variation. 

The main purpose of this paper is to evaluate the accuracy of the resulting dark-sky flats obtained using various observational techniques in a given condition. Thus, we used a constant value corresponding to an SB of $\sim$21.5 mag arcsec$^{-2}$ (see Section \ref{sec:obs:image}), rather than accounting for the complex nature of AG. The potential impacts of AG fluctuations are instead discussed using a simplified model in Section \ref{sec:disc:ag}. 

ZL consists of thermal radiation emitted from interplanetary dust and sunlight scattered by interplanetary dust. Although its brightness depends on solar activity and the Earth's relative position to the interplanetary dust plane, the temporal variation over a single night is likely insignificant. Instead, ZL creates a large-scale spatial gradient in the night sky, as its brightness varies smoothly with the angular distance from the Sun. The ZL map was created using the Python package \texttt{ZodiPy} \cite[][]{2024JOSS....9.6648S}, which implements the interplanetary dust model from \cite{1998ApJ...508...44K} and the reflectance spectrum in optical wavelength from \cite{2017PASJ...69...31K}. 

DGL is the faintest of the three components, except in regions near the Galactic plane. Nevertheless, its small-scale structures are likely to be confused with small, faint galaxies and also introduce spatial fluctuations that hinder accurate sky background subtraction, making it one of the most critical background components to control in LSB observations \cite[see][]{2019arXiv190909456M,2020A&A...644A..42R,2025ApJ...979..175L}. 

We calculated the DGL intensity at a wavelength ($I_\lambda$) using infrared data from the Infrared Astronomical Satellite \cite[IRAS; ][]{1984ApJ...278L...1N} and the Cosmic Background Explorer \cite[COBE; ][]{1992ApJ...397..420B}. When the dust optical depth ($\tau_\lambda$) is less than 1, the DGL intensity is assumed to be proportional to the 100-$\mu$m intensity ($I_\nu$):
\begin{equation}
\lambda I_\lambda = \alpha_\lambda(\nu I_\nu)_{100\mu m},
\end{equation}
where the proportional constant ($\alpha_\lambda$), known as the correlation spectrum, is determined by the product of the interstellar radiation field---comprising the integrated starlight of the Milky Way (MW)---and the reflectance of interstellar dust. However, when $\tau_\lambda > 1$, a correction factor ($\beta_\lambda$) is applied under the assumption that dust and stars are mixed homogeneously:
\begin{equation}
\lambda I_\lambda = \alpha_\lambda\beta_\lambda(\nu I_\nu)_{100\mu m}, 
\end{equation}
where $\beta_\lambda$ is given by
\begin{equation}
\beta_\lambda = \frac{1-\mathrm{exp}(-\tau_\lambda)}{\tau_\lambda}.
\end{equation}

The optical depth was computed by utilizing the $E(B-V)$ maps of \cite{1998ApJ...500..525S}, and its wavelength dependence was calculated using the extinction curve of \cite{1999PASP..111...63F}. Since the spatial resolution of the 100-$\mu$m map is 4$\arcmin$, we resampled the DGL map to match the pixel scale of the K-DRIFT data by upscaling it using interpolation. Meanwhile, we compared the correlation spectrums from various studies \cite[e.g.,][]{2012ApJ...744..129B,2015ApJ...806...69A,2017PASJ...69...31K,2022ApJ...932..112C} and found that their absolute values vary by up to a factor of two, but their wavelength dependence remains similar. Hence, we adopted the result of the most recent study \cite[][]{2022ApJ...932..112C}, which were obtained from the Baryon Oscillation Spectroscopic Survey \cite[][]{2013AJ....145...10D} data. 

Finally, we employed two contrasting sub-regions of the sky background for the case study: the ``clear'' and ``dirty'' skies (Figure \ref{fig:f1}). The primary difference between them is ``weak ZL and DGL'' and ``strong ZL and DGL.'' The clear sky is a region with minimal DGL, typically found at high galactic latitudes, chosen for its exceptionally weak ZL. While the absolute intensity of ZL is low, a brightness gradient of approximately 15\% spans from the top left to the bottom right of the frame. In contrast, the dirty sky is a region with strong DGL at low galactic latitude, where ZL is also strong; however, the most striking feature is the complex structure of DGL. 

It is worth mentioning that K-DRIFT primarily focuses on LSB studies and thus will rarely observe the so-called zone of avoidance---galactic latitudes within approximately $\pm$15$\arcdeg$---where the bulk of the extinction ($A_V$) exceeds 0.5. Nevertheless, by employing the dirty sky in this comparative experiment, we aim to demonstrate how effectively data quality can be improved even under the worst situations. 

\subsubsection{Mock Objects} \label{sec:sim:obj}
Dark-sky flats are generated by combining a set of science images---typically ranging from a few tens to several hundreds---depending on the observation design. Before this step, it is essential to mask light sources, such as celestial objects, cosmic rays, and satellite trails. Commonly-used tools like \texttt{Source Extractor} \cite[][]{1996A&AS..117..393B} are typically available for object masking. However, some light may remain unmasked due to the complex interplay between tool settings and data quality. This residual light can cause an error, as if the regions are intrinsically brighter, potentially compromising the accuracy of the dark-sky flat. To incorporate this issue into the error budget, we generated mock objects, including stars and galaxies, and superimposed them onto the simulated sky backgrounds. Note that these objects were added directly to the sky background without accounting for the extinction effect by DGL. 

\begin{figure}[t!]
\includegraphics[width=\linewidth]{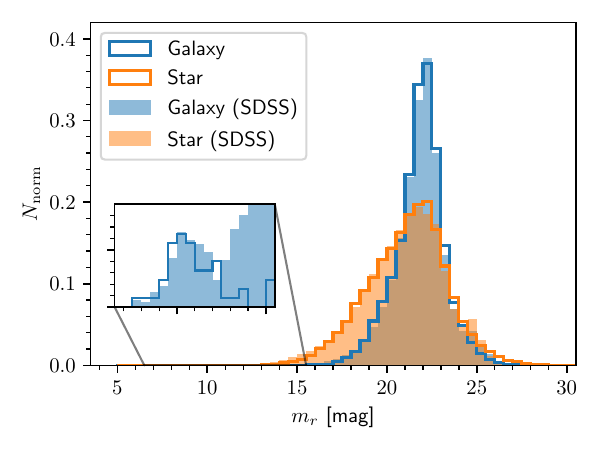}
\caption{Histograms of the $r$ magnitudes of galaxies (blue) and foreground stars (orange). The populations were derived from randomly selected subsamples of the SDSS catalog. The presence of some bright galaxies indicates the inclusion of nearby massive galaxies. \label{fig:f2}}
\end{figure}

We first estimated the number and magnitude of celestial objects within twelve arbitrarily selected subregions from the Sloan Digital Sky Survey (SDSS) data. Based on the results, we assumed that there were approximately one million foreground stars and 1.5 million galaxies within the $\sim$100 deg$^2$ area, which corresponds to the total area of the employed sky frames. Figure \ref{fig:f2} shows the magnitude distribution of the mock objects: almost all foreground stars and galaxies lie within $\sim$15--25 mag.\footnote{Considering the noise used in this experiment, the limiting magnitude of a single image is estimated to be $\sim$21 mag.} To mimic nearby massive galaxies, $\sim$1\% of mock galaxies were assigned to be brighter than 15 mag. 

Mock objects were generated using simple two-dimensional (2D) models from the Python package \texttt{astropy.modeling.models}, and the number density in each magnitude bin was adjusted to match the values provided earlier. For foreground stars, we employed a Moffat model defined by the following formula:
\begin{equation}
\mathrm{Moffat}\ I(r)=A\left(1+\frac{r^2}{\gamma^2}\right)^{-\alpha},
\end{equation}
where $A$ is an arbitrary amplitude factor, and $\gamma$ and $\alpha$ are seeing-dependent parameters. In this study, we fixed these parameters at 2.03 and 3.88, respectively, based on updated estimates from the upgraded K-DRIFT P1 observations, following the analysis by \cite{2022PASP..134h4101B}. 

For galaxies, a S\'ersic profile was adopted, described by the following function:
\begin{equation}
\mathrm{S\acute{e}rsic}\ I(r)=I_e \mathrm{exp}\left\{-b_n\left[\left(\frac{r}{r_e}\right)^{1/n}-1\right]\right\},
\end{equation}
where $r_e$ is the effective radius, $I_e$ is the intensity at the effective radius, $n$ is the S\'{e}rsic index. The constant $b_n$ is determined such that it satisfies $\Gamma(2n)=2\gamma(2n,b_n)$, where $\Gamma$ and $\gamma$ are the gamma and low incomplete gamma functions, respectively. For simplicity, we used the approximation $b_n=2n-1/3$. In practice, the parameters to be determined were total brightness and $r_e$, while other parameters were assigned randomly within given ranges: the S\'ersic index between 1 and 2, ellipticity from 0 to 0.9, and position angle (PA) from 0\arcdeg to 180\arcdeg. Since the galaxies would eventually be masked, we did not consider their complex structures, such as spirals and irregularity. 

Finally, we randomly placed over two million simulated objects across the simulated sky background. At this time, we applied seeing effects to the mock galaxies by convolving them with the same parameters used for the mock star model. Figure \ref{fig:fA-1} in Appendix \ref{sec:app:obj} shows the resulting mock object map, enhanced with some noise for better visibility. When generating the set of $r_e$ values, we intentionally used a probabilistic approach biased toward smaller sizes. As a result, the mock galaxies are seemingly located at various distances. Since we did not consider the real 3D distribution, clustering-like aspects are rarely seen in both stars and galaxies. 

\begin{figure*}[t!]
\includegraphics[width=\linewidth]{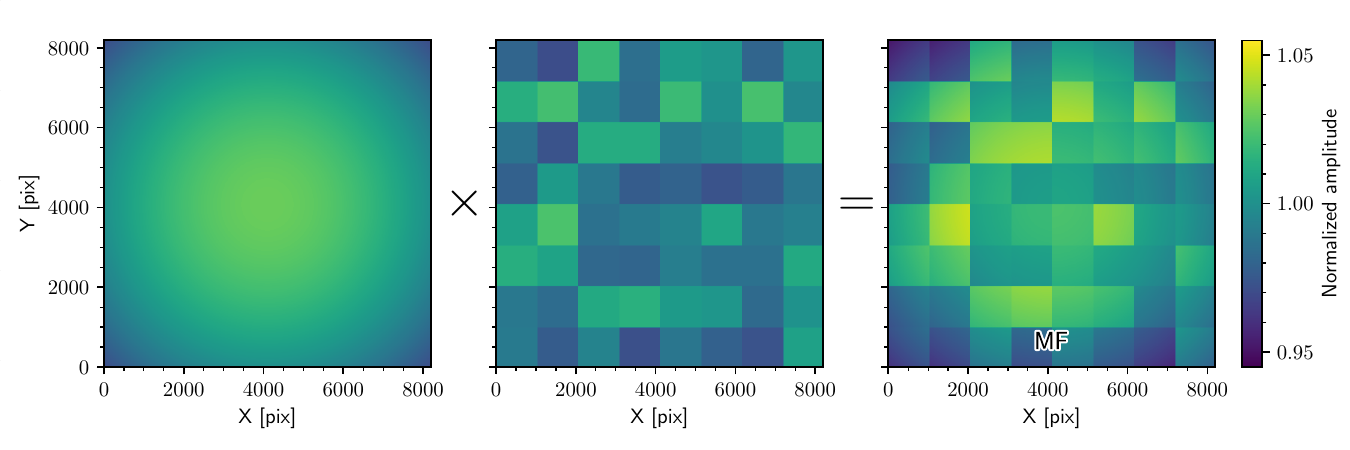}
\caption{Pattern of the mock flat, composed of a combination of a Gaussian kernel and a mosaic configuration. The peak-to-peak deviation of the mock flat is $\pm$5\%. \label{fig:f3}}
\end{figure*}

It is worth noting that crowded light sources from clustering can impact the robustness of masking results in practice. For example, the superposition of point spread function (PSF) effects in crowded regions may contaminate local sky-level estimates when masking objects.\footnote{A more detailed discussion on the impact of overcrowding and associated PSF effects on sky estimation and source detection can be found in \cite{2014A&A...567A..97S,2015A&A...577A.106S}, \cite{2016ApJ...823..123T} and \cite{2023MNRAS.520.2484K}.} Furthermore, as noted in Section \ref{sec:obs:dsf}, overcrowded light sources can lead to larger masked areas, which may result in blank pixels in dark-sky flats if the dithering size is too small. 

\subsection{Mock Observation} \label{sec:obs}
We designed the configuration of the mock data to resemble the dataset from the K-DRIFT survey. Each frame is an $\mathrm{8k}\times\mathrm{8k}$ image with a pixel scale of 2$\arcsec$, yielding an FoV of $\sim$$4.5\arcdeg\times4.5\arcdeg$. In typical observations, the sky background level changes continuously due to varying conditions, such as changes in airmass. However, in this experiment, we assumed the sky background remains constant as the observations are expected to be conducted near the zenith over a short duration. Further preparations are described in the following sections. 

\subsubsection{Mock Flat} \label{sec:obs:flat}
Before obtaining mock data, the pixel-to-pixel sensitivity variation, i.e., the mock flat (MF), was set in advance. While the sensitivity variation of the actual sensor might have complex structures depending on the manufacturing process, we instead employed an artificial pattern generated using some parametric functions. Figure \ref{fig:f3} illustrates the MF pattern generated by combining a center-symmetric Gaussian kernel with a mosaic configuration. The peak-to-peak deviations of both components are approximately $\pm$3\%, leading to an overall deviation of $\pm$5\% in the resulting MF. 

Note that the camera to be mounted on the K-DRIFT telescope consists of a single large CMOS chip, making it unlikely to exhibit discontinuities like the mosaic pattern used in this experiment. However, we intentionally imitated the mosaic CCD chip configurations of conventional wide-field observation systems, such as the Hyper Suprime-Cam \cite[][]{2018PASJ...70S...1M}, the Dark Energy Camera \cite[][]{2015AJ....150..150F}, and the Large Synoptic Survey Telescope Camera \cite[][]{2019ApJ...873..111I}. This noticeable pattern not only enhances the visibility of the dark-sky flats obtained in this study but also helps us evaluate their accuracy. 

\begin{figure*}[t!]
\includegraphics[width=\linewidth]{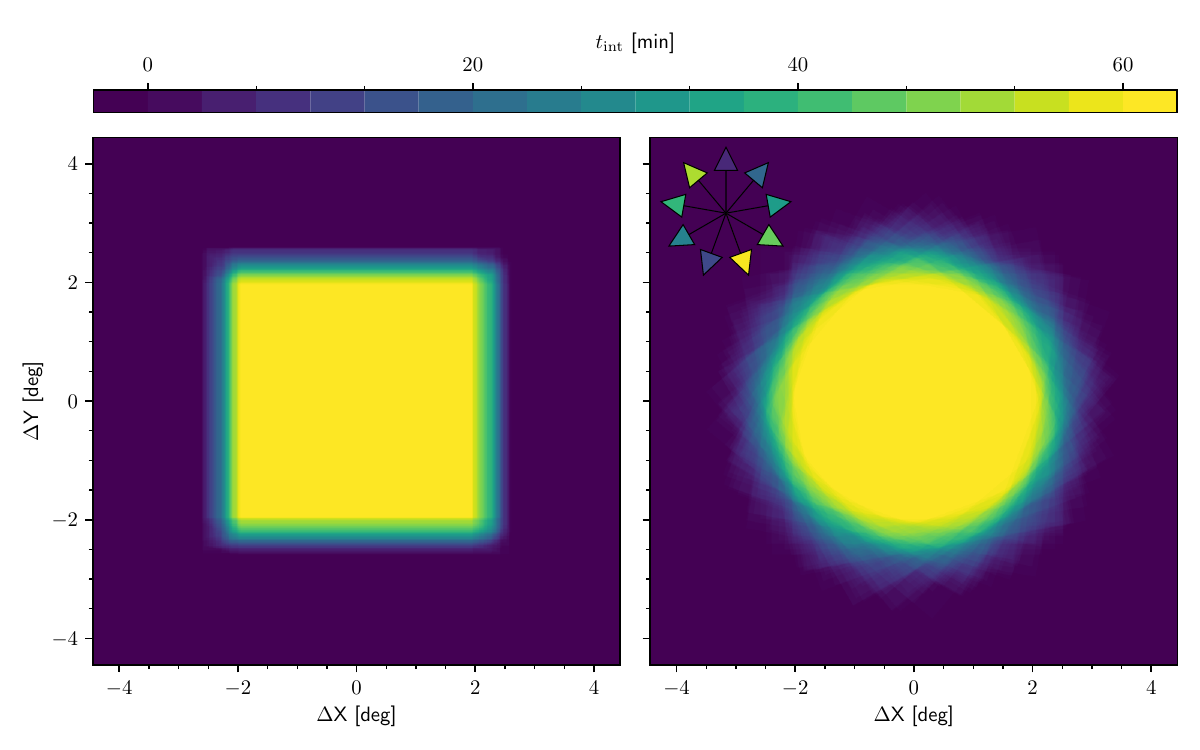}
\caption{Coverage of an observation sequence for the offset dithering method (left) and the rolling dithering method (right). The integration time at the end of the sequence is color-coded. For both methods, the $xy$-shift is randomly determined within a $40\arcmin\times40\arcmin$ box area relative to the reference coordinate. The direction and order in which the PA rotates during the rolling dithering method are displayed in the upper left of the right panel. \label{fig:f4}}
\end{figure*}

\subsubsection{Observation Methods} \label{sec:obs:dither}
Image combining---typically via median combining---is one of the most effective statistical approaches to reducing noise and flattening the sky background. When combined with dithering observations, it also helps prevent defective or masked pixels from overlapping between frames.\footnote{Sampling effects with subpixel dithering may improve the astrometric and photometric accuracy, thus mitigating the resolution limit of K-DRIFT.} When generating dark-sky flats, masked pixels in one frame are filled with pixels from other frames observing the sky background. In this context, the accuracy of the dark-sky flat depends heavily on the dithering methods, including the scale and complexity of the $xy$-shift pattern. 

From the early stages of K-DRIFT, we have considered incorporating a rotating camera---i.e., allowing the detector to rotate, thus varying its PA relative to the sky---as a method to reduce flat-fielding errors. Indeed, this concept, previously also employed by \cite{2016ApJ...823..123T}, has proven its effectiveness in LSB observations. To assess how significantly the camera rotation improves the accuracy of dark-sky flats, we implemented the following two dithering methods:
\begin{itemize}
\item{The ``offset'' dithering is a conventional observing pattern that involves applying $xy$-shifts via telescope mount between exposures. The $xy$-shifts are randomly determined within a $40\arcmin\times40\arcmin$ box area relative to the reference coordinate. This box size---set to be several times larger than the angular extent of MW if placed at a distance of 20 Mpc---reflects a typical case of observing a nearby, large galaxy, ensuring that masked pixels from these objects do not overlap between frames.}
\item{The ``rolling'' dithering is similar to offset dithering in that it also applies $xy$-shifts. However, after multiple exposures, the camera's PA is rotated by 160$\arcdeg$, and observations resume. After nine rotations, the camera returns to its original PA, completing an observation sequence. In summary, rolling dithering corresponds to conducting nine offset dithering observations at nine distinct PAs. It is worth noting that we use an interval of 160$\arcdeg$ instead of 40$\arcdeg$ to prioritize obtaining data at near-opposite PAs, even if the sequence ends prematurely during the actual observations.}
\end{itemize}

The number of frames used in this experiment was determined by the following considerations: First, the number of frames must be statistically sufficient to ensure reliable dark-sky flats. Second, the duration of the observation sequence must be kept relatively short, as the K-DRIFT observations will be conducted while the target is near the zenith. Third, the exposure time of each frame must be long enough to prevent LSB signals from being overwhelmed by shot noise. Taking these factors into account, we opted for a 1-hour sequence with a single exposure time of 40 sec, yielding a total of 90 frames.\footnote{In actual K-DRIFT observations, each target will likely be observed for more than two hours per night, spanning the period before and after culmination. This means that the one-hour observation sequence will be repeated multiple times.}

Figure \ref{fig:f4} shows the coverage of a single observation sequence for each dithering method. While the FoV of a single frame is $\sim$20 deg$^2$, the deepest area, where all frames overlap, is reduced to approximately 14.5 and 11.5 deg$^2$ for the offset and rolling dithering methods, respectively. The rolling dithering method will slightly decrease survey efficiency due to its circular coverage, which is smaller and less optimal for filling large survey areas. However, if it effectively mitigates flat-fielding errors, this compromise could ultimately enhance LSB research capabilities in individual regions. 

\subsubsection{Mock Image Acquisition} \label{sec:obs:image}
As mentioned in Section \ref{sec:sim:bkg}, we assumed that AG remains constant and that observations are conducted near the zenith, allowing us to ignore any changes in sky brightness over an hour. Instead, we incorporated random noise into each frame. The process for generating mock images is as follows, repeated 90 times for each dataset: 
\begin{enumerate}
\item{Prepare the simulated night sky and MF frames: The SB of the night sky ($\mathrm{ZL}+\mathrm{DGL}$) is converted to count using the equation $\log \mathrm{count} = 0.4\times(m_0-\mu)$, where $m_0$ is the photometric zero point assumed to be 30.5, and $\mu$ is the SB in mag arcsec$^{-2}$.}
\item{(For rolling dithering only) Rotate the night sky frame: After every ten iterations, the frame is rotated by 160$\arcdeg$ about the center. The rotation is implemented using the Python package \texttt{scipy.ndimage.rotate} with a spline interpolation order of 0.}
\item{Determine a pointing coordinate for exposure: A random $xy$-shift relative to the reference coordinate is applied. The reference coordinate is set at the center of the night sky frame.}
\item{Extract a subframe: An $\mathrm{8k}\times\mathrm{8k}$ subframe centered on the pointing coordinate is extracted from the night sky frame.}
\item{Add an AG component: A constant value corresponding to an SB of 21.5 mag arcsec$^{-2}$ is incorporated into the subframe. The conversion follows the same equation used in Step 1.}
\item{Introduce background noise: 2D random Gaussian noise is incorporated into the subframe, with a mean of zero and a standard deviation corresponding to 10\% of the total sky background level.}
\item{Apply the detector's illumination pattern: The subframe is multiplied by the MF frame to account for pixel-to-pixel sensitivity variations.}
\item{Save the resulting subframe as a FITS file, with the saturation level set to 65,535 counts, reflecting the typical limit of 16-bit data.}
\end{enumerate}

In conclusion, we obtained four datasets by employing two distinct sky backgrounds (clear and dirty), two dithering methods (offset and rolling), and one MF. The mean values of the sky backgrounds for these datasets are approximately 4650 and 6380 counts for the clear and dirty skies, respectively. 

\subsubsection{Dark-Sky Flat Generation} \label{sec:obs:dsf}
To obtain reliable dark-sky flats, it is essential to mask all light sources except the sky background before combining science images. We used \texttt{Source Extractor} for systematic masking with an initial detection threshold of 1.5$\sigma$ above the sky background level. As expected, relatively compact and bright objects were effectively masked, but some extended diffuse light remained unmasked. To mask the residual light, we manually applied circular masks with diameters twice the major-axis length of each initially masked segment. Note that the size of circular masks required to properly exclude extended light may vary depending on the initial detection threshold, image quality, and source crowdedness in real observations. 

\begin{figure}[t!]
\includegraphics[width=\linewidth]{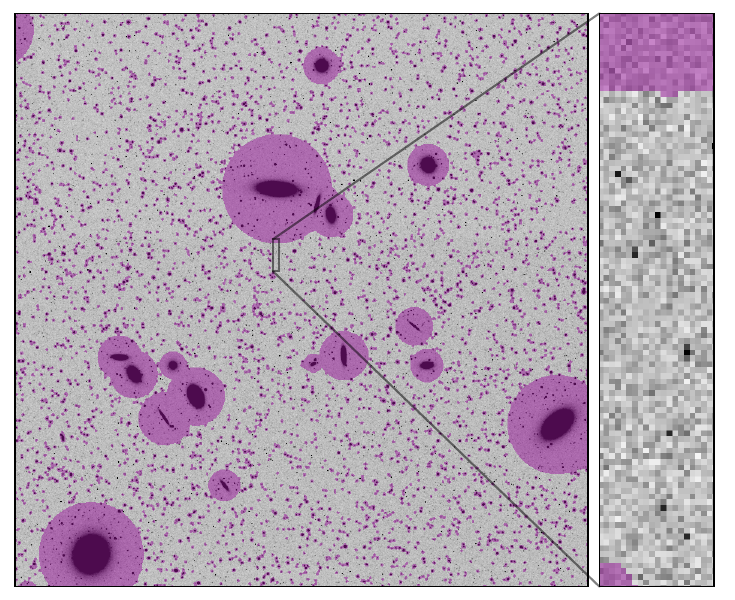}
\caption{Example of masking results showing a subregion with an FoV of $1\arcdeg\times1\arcdeg$, extracted from an arbitrary mock image. The masked regions are highlighted in magenta. Several circular masks cover the extended light surrounding large and/or bright objects. The zoomed-in image shows some unmasked signals remaining in the background. \label{fig:f5}}
\end{figure}

Figure \ref{fig:f5} shows an example of the masking results for a subregion with an FoV of $1\arcdeg\times1\arcdeg$. In each mock image, approximately 15\% of the area is masked. Although more aggressive masking processes, such as increasing the diameter of manual masks, could be implemented, it is important to keep in mind that overlapping masked regions may lead to blank pixels in dark-sky flats. 

Meanwhile, some signals from tiny objects or hot pixels introduced by noise generation remained unmasked because the minimum segment size was set to five pixels. However, these pixel-scale signals are expected to be eliminated during image combining, as they will likely be overwhelmed by scatter or excluded as outliers in median calculations. Even if such signals introduce pixel-scale fluctuations, they are unlikely to significantly impact the results for the large-scale fluctuations that are the primary focus of this study. 

Finally, we median-combined the object-masked mock images from each dataset. Although the night skies employed in this experiment remain constant over time, dithering introduces minor offsets in sky brightness between images. To address this, we normalized each dataset during image combining by scaling each image to its respective sky background level. Note that dark-sky flats may vary slightly depending on the specific combining methods, such as the type of calculation or scaling factor; however, these details are beyond the scope of this paper. 

\section{Results} \label{sec:result}
This section compares and describes the details of the four dark-sky flats resulting from different sky backgrounds and observation methods. For clarity, we refer to the four dark-sky flats as ``coDSF,'' ``crDSF,'' ``doDSF,'' and ``drDSF,'' using the abbreviations for the setups, outlined in Table \ref{tab:t1}. 

\begin{deluxetable}{ccccc}
\tabletypesize{\small}
\tablenum{1}
\tablecaption{Summary of four dark-sky flats. \label{tab:t1}}
\tablewidth{0pt}
\tablehead{
\multirow{2}{*}{Name} & \multirow{2}{*}{\shortstack{Sky \\ background}} & \multirow{2}{*}{\shortstack{Dithering \\ method}} & \multirow{2}{*}{$N_\mathrm{stack}$} & \multirow{2}{*}{$C_\mathrm{bkg}$} \\
}
\startdata
coDSF & Clear & Offset & 90 & 4634 \\
crDSF & Clear & Rolling & 90 & 4634 \\
doDSF & Dirty & Offset & 90 & 6350 \\
drDSF & Dirty & Rolling & 90 & 6357 \\
\enddata
\tablecomments{$N_\mathrm{stack}$ is the number of coadded frames used to generate the dark-sky flat. $C_\mathrm{bkg}$ refers to the count that yields the mean of the residuals to be zero, not the mean of the dark-sky flat itself.}
\end{deluxetable}

\begin{figure}[t!]
\includegraphics[width=\linewidth]{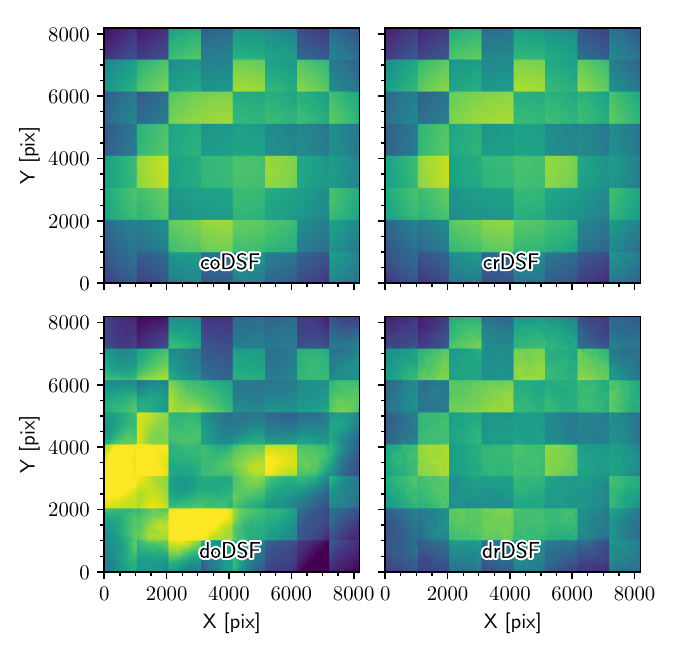}
\caption{2D maps of the four resulting dark-sky flats. Each map is labeled with its name, and the color scales are the same as Figure \ref{fig:f3}. All the maps appear comparable, except for doDSF, which shows significant fluctuations. \label{fig:f6}}
\end{figure}

Figure \ref{fig:f6} presents the 2D maps of the four dark-sky flats. The most noticeable result is doDSF, which exhibits a severe abnormal excess of light. Masking alone does not completely remove DGL from the mock images, leaving some fluctuations. In contrast, coDSF, crDSF, and drDSF are nearly indistinguishable from one another and closely resemble MF. Interestingly, drDSF shows much weaker fluctuations despite using the same masking operation and sky background as doDSF. 

This preliminary result indicates that the rolling dithering method can generate more accurate dark-sky flats by mitigating signals that cannot be adequately eliminated through masking. However, since this visual assessment can be somewhat naive, we present residual maps below to facilitate a more detailed and quantitative comparative analysis. 

\subsection{Residual Maps} \label{sec:result:resi}
We computed the residual maps ($\epsilon$) using the following definition:
\begin{equation}
\epsilon_\mathrm{xDSF} \equiv \mathrm{MF} - \frac{\mathrm{xDSF}}{C_\mathrm{bkg}},
\end{equation}
where $\mathrm{x \in \{co,\ cr,\ do,\ dr\}}$. To account for the different levels of the dark-sky flats depending on the experimental setup, we normalized each by dividing it by its corresponding background level ($C_\mathrm{bkg}$). The $C_\mathrm{bkg}$ values were determined through a parameter search, defined as the integers that minimize the absolute mean of the residuals to less than 10$^{-4}$; these values are noted in Table \ref{tab:t1}. 

\begin{figure*}[t!]
\includegraphics[width=\linewidth]{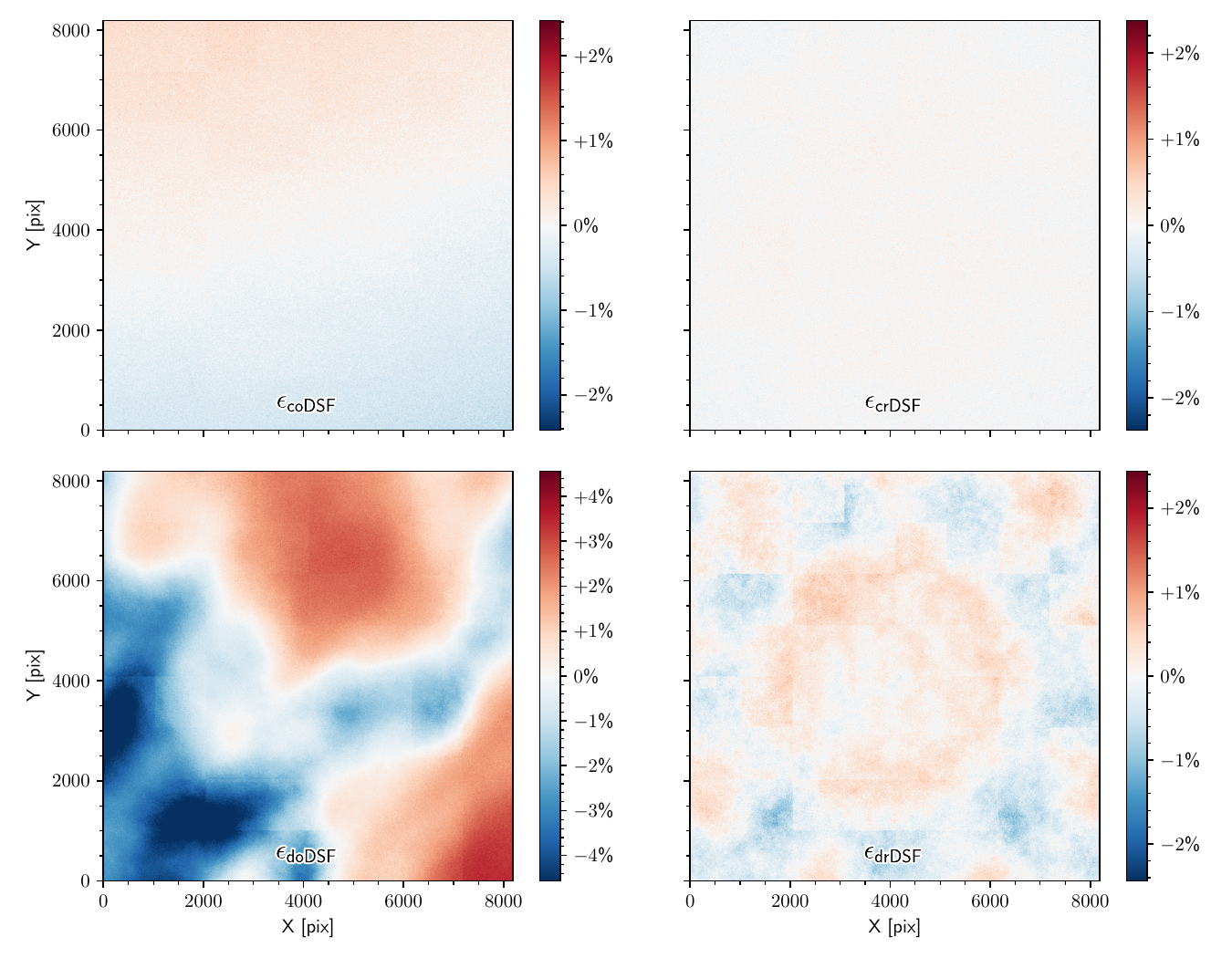}
\caption{Residual maps obtained by subtracting each dark-sky flat from MF. The upper panels show the results for the clear sky using the offset dithering (left) and rolling dithering (right) methods. The lower panels show the results for the dirty sky, also using the offset dithering (left) and rolling dithering (right) methods. The color scales are limited to the $\pm$2$\sigma$ range in each case. \label{fig:f7}}
\end{figure*}

Figure \ref{fig:f7} shows the residual maps obtained by subtracting the dark-sky flats from MF. Positive (negative) values indicate that flat-field correction using the corresponding dark-sky flats leads to over- (under-) correction, amplifying (diminishing) the pixel brightness. To enhance visibility by suppressing outliers caused by random noise, the color scales are limited to the $\pm$2$\sigma$ range. The detailed analyses are as follows:
\begin{itemize}
\item{$\epsilon_\mathrm{coDSF}$: The standard deviation across the full frame is $\sim$1\%, and the spatial peak-to-peak deviation\footnote{This is calculated as the difference between the maximum and minimum of the median values from a random sampling of 5000 sub-regions, each measuring $100\times100$ pixels.} is $\pm$0.5\%. The most prominent feature is a gradient caused by the spatial variation of ZL. Its brightness varies smoothly from the top left to the bottom right of the frame. Since coDSF contains a ZL-induced gradient that is not present in MF, flat-field corrections using coDSF can seemingly flatten the mock images. However, in practice, this approach introduces photometric errors of up to 1\% for each image. The magnitude of these errors is strongly correlated with the FoV and will pose a significant problem as the FoV increases. The mosaic pattern is barely visible, with brightness differences at the mosaic boundaries remaining below 0.05\%.}
\item{$\epsilon_\mathrm{crDSF}$: The standard deviation is comparable to that of coDSF. However, the gradient disappears because the ZL effect is averaged out as the frame rotates, resulting in a spatial peak-to-peak deviation of $\pm$0.1\%. Indeed, subtracting the gradient from coDSF through 2D modeling produces a result nearly identical to crDSF. The mosaic pattern is almost unrecognizable, with brightness differences similar to that of coDSF. Consequently, employing the rolling dithering method under clear-sky conditions ensures the generation of a robust dark-sky flat. Flat-field corrections using crDSF address only the detector's sensitivity variations while leaving the intact ZL component in each image. The remaining ZL can be corrected later through sky subtraction using 2D modeling.}
\item{$\epsilon_\mathrm{doDSF}$: The standard deviation is $\sim$2\%, and the spatial peak-to-peak deviation reaches up to $\pm$4.5\%. Despite the $xy$-shift, DGL is partially superimposed on the same pixels, leaving noticeable fluctuations in doDSF. Moreover, due to the large amplitude of these fluctuations, the mosaic pattern becomes more evident. If these features persist, flat-field correction using doDSF leads to severe local fluctuations rather than addressing the detector's sensitivity variations. In addition to introducing significant photometric errors in each image, the complex local fluctuations make it challenging to flatten the image, even with dedicated sky subtraction. This outcome is not particularly surprising, as it was initially expected to represent the worst-case scenario.}
\item{$\epsilon_\mathrm{drDSF}$: Despite using the dirty sky, the standard deviation is comparable to those obtained with the clear sky. If considering only the standard deviation, this suggests that the rolling dithering method plays a significant role in improving the accuracy of dark-sky flats. However, nearly point-symmetric fluctuations remain, with a spatial peak-to-peak deviation of $\pm$1\%. Ideally, DGL is expected to be azimuthally averaged out as the camera rotates, but in practice, the median of each pixel is not correctly calculated due to DGL overlap. For this reason, the shape of these fluctuations can vary depending on the mask size and the dithering interval (see details in Section \ref{sec:disc:bs} and \ref{sec:disc:pa}). As a result of these fluctuations, the mosaic pattern itself becomes more prominent, although the deviated values are relatively small.}
\end{itemize}

Consequently, we propose that the optimal approach to obtaining an accurate dark-sky flat is to observe a clear sky with weak ZL and DGL signals employing the rolling dithering method. This approach enables accurate sky subtraction and estimation, supporting robust LSB studies. For this reason, the K-DRIFT survey is primarily designed to target skies with minimal ZL and low DGL on any given day. 

Meanwhile, the rolling dithering method would still be effective even if ZL is relatively strong. However, when DGL is strong, using the alternative would be more adequate. Assuming that the camera's performance has not changed significantly, dark-sky flats obtained under clearer conditions on the nearest possible day can serve as the master flat. In the worst-case scenario, where no alternatives are available, the rolling dithering method would be preferable to conventional methods. 

\subsection{Impact on Photometric Quality and Subsequent Consequence} \label{sec:result:impact}
In LSB research, photometric quality is often characterized by the detectable surface-brightness limit, typically estimated from the standard deviation of the sky background. This value can vary depending on how the measurement area is defined and can also be standardized based on the characteristics of the object of interest \cite[see Appendix A of ][]{2020A&A...644A..42R}. 

In the previous section, we found that the standard deviations for the three results ($\epsilon_\mathrm{coDSF}$, $\epsilon_\mathrm{crDSF}$, and $\epsilon_\mathrm{drDSF}$) were $\sim$1\% across the full frame. If photometric errors arise solely from flat-field correction, one might expect the surface-brightness limits, estimated from the background noise of the outcomes corrected with these dark-sky flats, to be comparable. However, the subsequent sky subtraction on each science image would be implemented influenced by their different underlying illumination patterns, thus affecting photometric quality in a cascading manner. 

To assess the subsequent impacts of differences in flat-field correction, we compared the final products after completing sky subtraction and coaddition. Due to the poorest flat-field correction, the accuracy of the outcome with doDSF is expected to be the worst, followed by drDSF among the other three. Consequently, we focused on comparing two coadded images obtained with coDSF and crDSF. 

Initially, flat-field correction was applied to each dataset using the respective dark-sky flats. Following this, sky subtraction was performed by modeling the sky background of each mock image with a second-order 2D polynomial fit. The object-masked mock images were divided into $16\times16$ grids, and a 2D background model was created using the sigma-clipped median value from each grid. The conceptual workflow ($\mathbb{F}$) of data processing for the two datasets can be schematically expressed as follows:
\begin{equation}
\mathbb{F}_\mathrm{coDSF} = \frac{\mathrm{Image}}{\mathrm{Flat}_G} - \mathrm{Sky}_{\bar{G}},
\end{equation}
\begin{equation}
\mathbb{F}_\mathrm{crDSF} = \frac{\mathrm{Image}}{\mathrm{Flat}_{\bar{G}}} - \mathrm{Sky}_G,
\end{equation}
where $G$ and $\bar{G}$ indicate ``with gradient'' and ``without gradient,'' respectively. Finally, we aligned each resulting dataset using the Python package \texttt{astroalign} \cite[][]{2020A&C....3200384B} and coadded them separately using \texttt{SWarp} \cite[][]{2002ASPC..281..228B}. Because the two coadded images have different coverages, we sampled the common region with an FoV of $\sim$$2.5\arcdeg\times2.5\arcdeg$ each for comparison. 

We found that the coadded image from $\mathbb{F}_\mathrm{coDSF}$ exhibits a standard deviation approximately 10\% larger than that of the image from $\mathbb{F}_\mathrm{crDSF}$, corresponding to a difference in the surface-brightness limit of $\sim$0.1 mag arcsec$^{-2}$. This discrepancy is relatively minor and does not strongly support the argument that flat-field corrections using dark-sky flats obtained through the conventional dithering method significantly degrade photometric quality. However, we should consider this error a lower limit as it was derived under clear-sky conditions. 

\begin{figure}[t!]
\includegraphics[width=\linewidth]{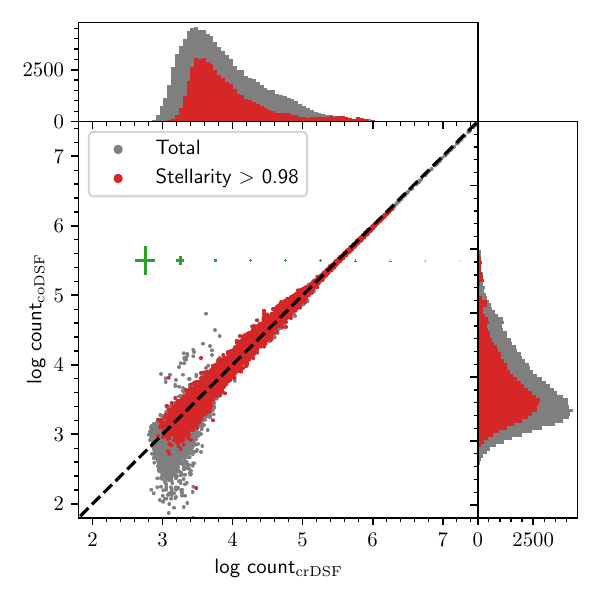}
\caption{Comparison of the source fluxes estimated from the two coadded images from $\mathbb{F}_\mathrm{coDSF}$ and $\mathbb{F}_\mathrm{crDSF}$. The gray and red data represent all detected sources and point-like sources, respectively. The dashed black line indicates a one-to-one relation, and the green bars show the mean errors for each bin. \label{fig:f8}}
\end{figure}

Figure \ref{fig:f8} compares the source fluxes estimated from the two coadded images. We used \texttt{Source Extractor} for the flux estimation and adopted FLUX\_BEST obtained with a detection threshold of 1.5$\sigma$. To ensure a fair comparison of fluxes for the same sources, we applied the double image mode with the coadded image from $\mathbb{F}_\mathrm{crDSF}$ as the reference. The flux discrepancies are pronounced for relatively faint sources (i.e., $\mathrm{log\ count \lesssim 5}$), likely due to the impact of the underlying gradient in coDSF. 

We then filtered the data to include only point-like sources with a stellarity greater than 0.98, where stellarity---defined as the CLASS\_STAR parameter from \texttt{Source Extractor}---is a value between 0 (galaxy-like) and 1 (star-like), based on source morphology and concentration. Most of the discrepancies observed in the faintest, extended sources disappear, aligning closely with a one-to-one relation. This result highlights two key points: First, extended sources are more susceptible to photometric errors caused by less accurate flat-field correction. Second, while the photometric zero points calculated using field stars remain broadly consistent between the two datasets, the uncertainty relative to $\mathbb{F}_\mathrm{coDSF}$ is expected to be more significant. 

In summary, this comparative analysis demonstrates that flat-field correction (and subsequent sky subtraction) can significantly impact photometric quality. These findings emphasize the critical importance of obtaining accurate dark-sky flats. The rolling dithering method plays a key role in achieving this accuracy, supporting the idea that the K-DRIFT survey based on this observation strategy holds promise for advanced LSB research. 

\begin{deluxetable}{clc}
\tabletypesize{\footnotesize}
\tablenum{2}
\tablecaption{Summary of experimental prescriptions considering various scenarios that may affect the accuracy of dark-sky flats. \label{tab:t2}}
\tablewidth{0pt}
\tablehead{
\colhead{Suffix} & \multicolumn{2}{c}{Description}
}
\startdata
\textbf{wMF5} & \multirow{4}{*}{\makecell[l]{The peak-to-peak deviation \\ of the mock flat is:}} & $\pm$5\%. \\
wMF7 & & $\pm$7\%. \\
wMF9 & & $\pm$9\%. \\
wMF11 & & $\pm$11\%. \\
\hline
\textbf{wAG0} & \multicolumn{2}{c}{The airglow does not fluctuate over time.} \\
wAG3 & \multirow{3}{*}{\makecell[l]{The peak-to-peak deviation \\ of fluctuating airglow is:}} & $\pm$3\%. \\
wAG6 & & $\pm$6\%. \\
wAG9 & & $\pm$9\%. \\
\hline
\textbf{woBS} & \multicolumn{2}{c}{No bright star is present within the FoV.} \\
wBS2 & \multirow{3}{*}{\makecell[l]{The bright star is masked \\ by a circle with a radius:}} & 2 times larger. \\
wBS4 & & 4 times larger. \\
wBS6 & & 6 times larger. \\
\hline
\textbf{wRoll9} & \multirow{4}{*}{\makecell[l]{One observation sequence \\ consists of a total of:}} & 9 PAs ($40\arcdeg\times n$). \\
wRoll5 & & 5 PAs ($72\arcdeg\times n$). \\
wRoll3 & & 3 PAs ($120\arcdeg\times n$). \\
wRoll2 & & 2 PAs ($180\arcdeg\times n$). \\
\enddata
\tablecomments{The four prescriptions indicated by boldface are the default setup of the main results in this study.}
\end{deluxetable}

\begin{figure*}[t!]
\includegraphics[width=\linewidth]{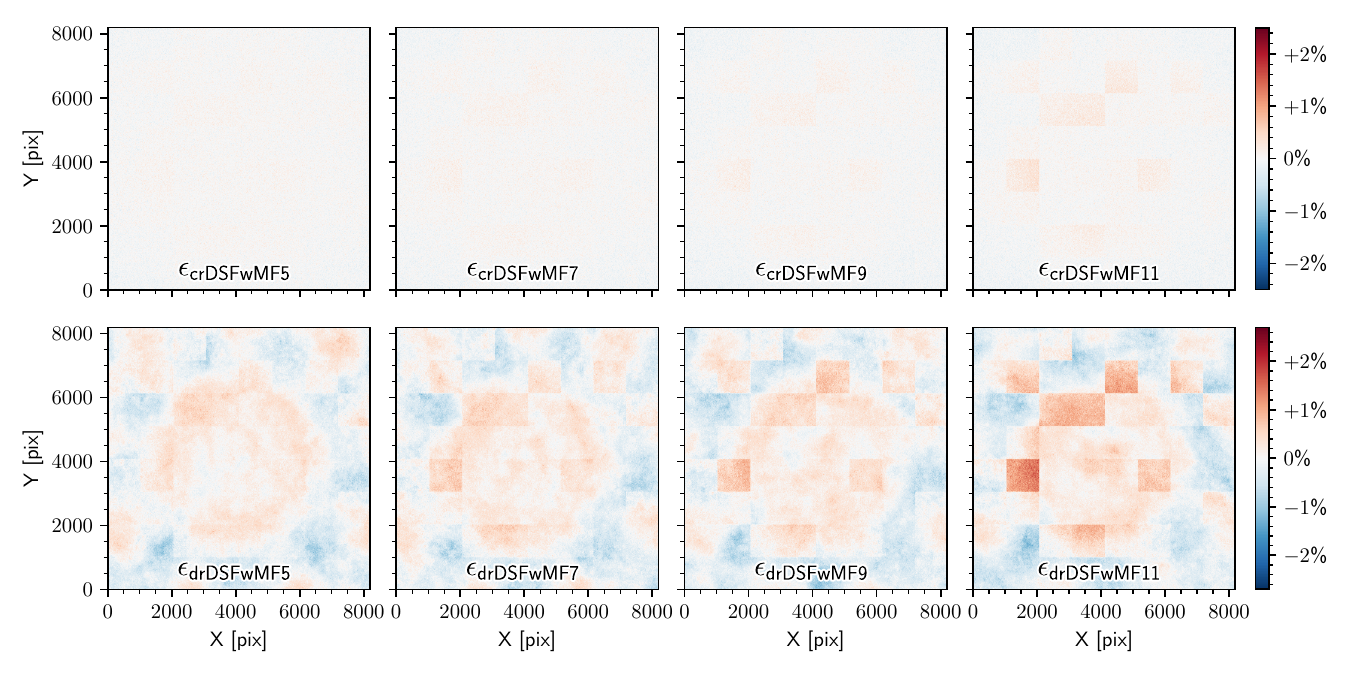}
\caption{Residual maps corresponding to different MF deviations. From left to right, the maps represent results for MF deviations of $\pm$5\%, $\pm$7\%, $\pm$9\%, and $\pm$11\%. The upper and lower panels show the results for the clear and dirty skies, respectively. The color scales are limited to the $\pm2\sigma$ range of the worst result for each sky condition. Detailed descriptions of each setup are provided in Table \ref{tab:t2}. \label{fig:f9}}
\end{figure*}

\section{Discussion} \label{sec:disc}
The findings thus far strongly indicate that the rolling dithering method is better suited for LSB studies, particularly in providing accurate flat-field correction. However, this conclusion is drawn from a single example based on a specific prescription, among many possibilities. In the following sections, we explore additional results derived from various prescriptions and discuss the effectiveness of the rolling dithering method more comprehensively. 

The standard deviation of the residuals is primarily determined by the random Gaussian noise introduced into each mock image, which theoretically decreases by a factor of $\sqrt{N}$, where $N$ is the total number of combined images. Thus, changing the number of images would yield highly predictable results, making it less interesting. Instead, we focused on other factors that could affect the imaging quality of individual science images. Given the vast parameter space, we concentrated on assessing four specific factors individually: (1) MF deviations, (2) AG fluctuations, (3) bright star and masking, and (4) different rolling intervals. Table \ref{tab:t2} summarizes the experimental prescriptions used in these assessments. 

\subsection{MF deviations} \label{sec:disc:mf}
In the main experiments, the MF deviation was constrained within $\pm$5\%, below the background noise level of 10\%. For this reason, its impact on generating dark-sky flats---specifically, calculating the median value of each pixel---was likely minimal. The rolling dithering method further enhances the accuracy of median calculation by allowing each pixel to sample a broader range of sky regions. This would have resulted in the mosaic pattern no longer noticeable in the residual maps. However, if the MF deviation becomes significant, how might it affect the accuracy of the dark-sky flats? 

To examine the impact of larger MF deviations on the accuracy of dark-sky flats generated using the rolling dithering method, we performed mock observations again with three new MFs, each having peak-to-peak deviations of $\pm$7\%, $\pm$9\%, and $\pm$11\%. These new MFs were generated by amplifying the deviations while preserving the overall shape of the original. Although modern sensors are unlikely to exhibit such significant pixel-to-pixel sensitivity variations, we intentionally considered these extreme cases to assess how increased deviations could affect the results. 

Figure \ref{fig:f9} shows the residual maps between the dark-sky flats and their respective MFs. The definition of the dark-sky flats is the same as in Figure \ref{fig:f7}, where a suffix is added to indicate differences in prescription conditions. Because the MFs have the same shape, the overall residual pattern does not change significantly. Additionally, the 2$\sigma$ limit remains nearly constant regardless of increases in the MF deviation. However, as the MF deviation increases, several blocks exhibit slightly high values. For example, the median value of the subregion $[x,\ y]$ = [1024:2048, 3072:4096] increases from approximately 0\% to 0.2\% for $\epsilon_\mathrm{crDSF}$ and from 0.3\% to 1\% for $\epsilon_\mathrm{drDSF}$. 

Upon examining the cause of this problem, we found that masking influenced the calculation of the median value. More specifically, as the MF deviation increases, both the signal and noise levels increase in regions where the MF value exceeds 1. This leads to excessive masking, even for pixels that have observed the sky background. Consequently, the median is underestimated because values from brighter regions are excluded, leaving only those observed in less bright regions. Through further experiments, we confirmed that this issue can be effectively mitigated by dividing the image into blocks based on the mosaic configuration and applying masking locally within each block. 

\begin{figure*}[t!]
\includegraphics[width=\linewidth]{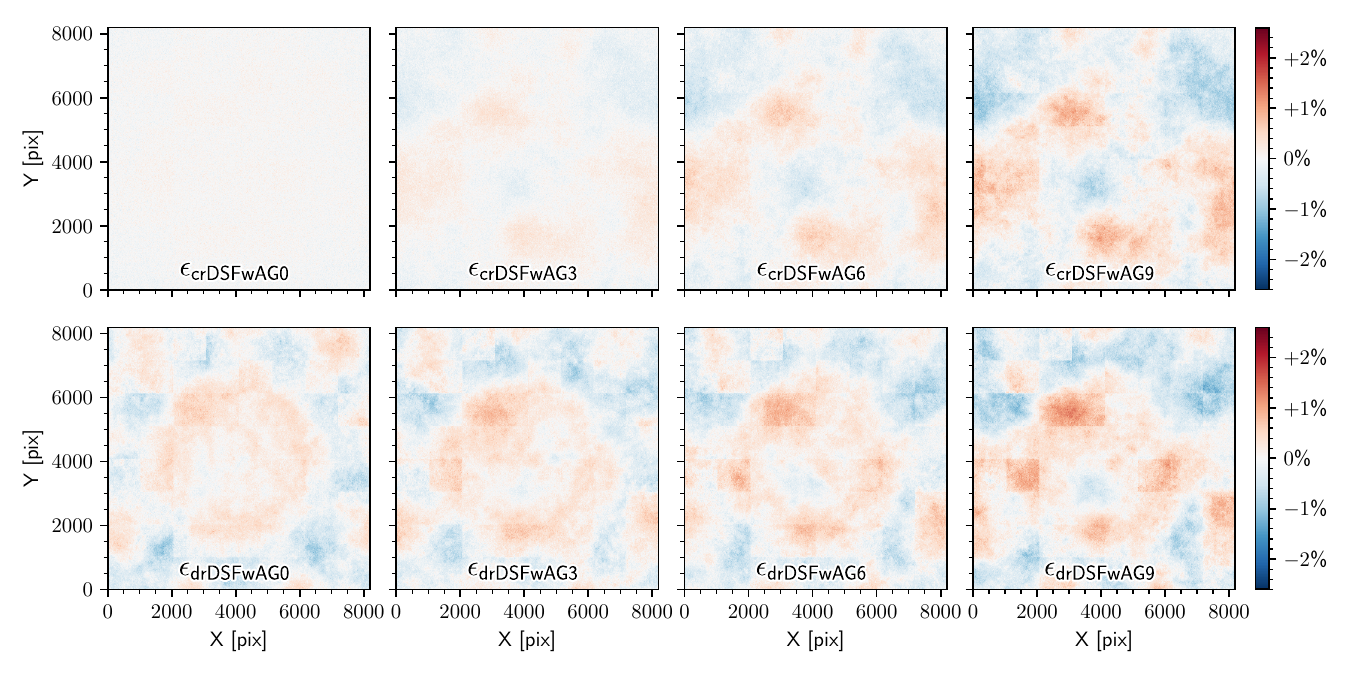}
\caption{Residual maps corresponding to different AG fluctuations. From left to right, the maps represent results for peak-to-peak deviations in AG brightness of 0\%, $\pm$3\%, $\pm$6\%, and $\pm$9\%. The upper and lower panels show the results for the clear and dirty skies, respectively. The color scales are limited to the $\pm2\sigma$ range of the worst result for each sky. Detailed descriptions of each setup are provided in Table \ref{tab:t2}. \label{fig:f10}}
\end{figure*}

In conclusion, the accuracy of dark-sky flats is closely related to the relative dominance of MF deviation and background noise. The experimental results suggest that masking must be applied more rigorously when the MF deviation is comparable to or exceeds the background noise level; this approach ensures minimal differences in the results. These findings confirm that the detector's pixel-to-pixel sensitivity variations do not pose substantial challenges to obtaining accurate dark-sky flats using the rolling dithering method. A more practical implication is that the masking strategy should effectively exclude celestial objects while preserving as many sky background pixels as possible in each frame, especially when the relative contribution of sensor variation is uncertain. 

\subsection{AG fluctuations} \label{sec:disc:ag}
AG emissions arise from various molecules at different atmospheric altitudes, resulting in highly complex fluctuations, even over short periods of time.\footnote{Examples of fluctuating AG can be found at the following website \url{https://skrutskie.uvacreate.virginia.edu/airglow/airglow.html}.} If AG fluctuates temporally in a random manner, its impact on the accuracy of dark-sky flats could be minimized through median combining. However, this cannot be guaranteed; it raises the question of what role the rolling dithering method might play, particularly if these fluctuations are not random. 

Simulating the intricate behavior of AG fluctuations is challenging and beyond the scope of this paper. To simplify the AG model, we adopted the following assumptions: First, cases involving sudden and explosive solar activities, such as geomagnetic storms, were excluded. Second, the overall AG brightness change was disregarded, as the observation sequence was assumed to be conducted for about an hour near the zenith. Consequently, our analysis focuses solely on local AG fluctuations over time. 

In signal processing, various types of noise are characterized by their spectral behavior. Among them, we employed Brownian noise, which is characterized by a spectral density following $S\propto1/f^\beta$ with $\beta=2$, to simulate large-scale fluctuations in AG. We first generated a 3D cube dataset of Brownian noise using a fixed random seed. Each layer along the $z$-axis, referred to as the $(i)$-th layer, was derived from the previous layer, the $(i-1)$-th layer. This approach provided similarity between adjacent layers, allowing the $z$-axis to approximate gradual changes over time. From this 3D dataset, we extracted nine sets of 10 slices at uniform intervals along the $z$-axis, where each set maintains continuity among its slices, but discontinuities exist between adjacent sets. These slices represent the time-varying AG during the observation sequence. Snapshots of the AG models are shown in Figure \ref{fig:fA-2} in Appendix \ref{sec:app:brown}. 

Finally, we re-performed mock observations by applying each slice in Step 5 of Section \ref{sec:obs:image}. To examine the impact of various AG fluctuation levels, we considered three scenarios with peak-to-peak deviations in AG brightness of $\pm$3\%, $\pm$6\%, and $\pm$9\%. These scenarios were created by amplifying the deviations while preserving the original fluctuation patterns. 

Figure \ref{fig:f10} presents the residual maps between the dark-sky flats and MF. While the 2$\sigma$ limit remains largely unchanged, local fluctuations become more prominent as the AG deviation increases. This impact is particularly noticeable under clear-sky conditions, leaving clumpy structures. In contrast, under dirty-sky conditions, local fluctuations caused by AG appear relatively insignificant because the residuals from DGL are already dominant. 

We naively expected that AG fluctuations would eventually average out, resulting in minimal impact on the dark-sky flats. However, this was not the case in practice, as the 90 noise patterns were not completely random. Given our experimental setup, similar noise patterns could persist up to 10 out of the 90 images, leading to local fluctuations; this impact resembles that of DGL-induced residuals. 

This result indicates that the persistence of fluctuating AG has a significant impact on the accuracy of dark-sky flats. Given the uncertainty in the true nature of AG, we must consider two possible scenarios. First, if AG fluctuations are more random, their impact is minimized, ensuring relatively accurate dark-sky flats. Indeed, we found that weakening the correlation between AG slices reduces the peak-to-peak deviation by half. Second, if AG fluctuations are highly persistent, the rolling dithering method could introduce additional randomness, improving the accuracy of dark-sky flats. Therefore, the rolling dithering method may still be effective in mitigating deviations in dark-sky flats. 

Another point worth mentioning is that the presence of AG fluctuations not only affects flat-field correction but also complicates the subsequent task of sky subtraction. Under clear-sky conditions without AG fluctuations, 2D plane modeling with a second-order polynomial fit enables straightforward and accurate sky subtraction. While higher-order polynomial functions may effectively remove AG fluctuations, they also risk eliminating LSB structures, such as diffuse galaxies and the outskirts of galaxies. To mitigate this challenge, we plan to conduct further experimental studies to improve sky subtraction techniques in future work. 

\begin{figure}[t!]
\includegraphics[width=\linewidth]{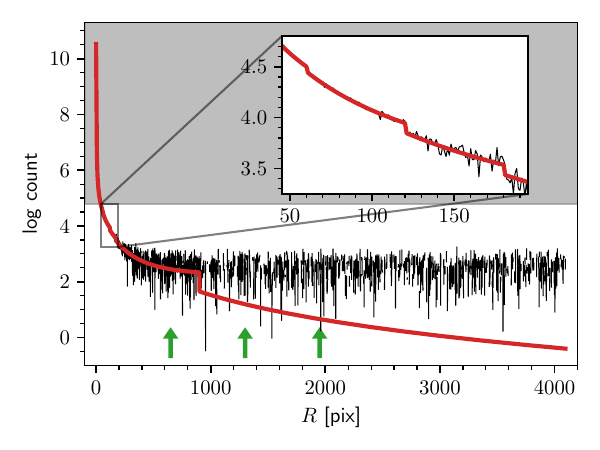}
\caption{Radial 1D profiles of a mock bright star. The solid red and black lines represent the original profile and the profile contaminated by background noise, respectively. The gray-shaded region indicates the saturated regime. The zoomed-in panel highlights the inner profile, where three of the four ghost features introduce subtle variations. The green arrows indicate the approximate radii of the three manual masks used in the experiment. \label{fig:f11}}
\end{figure}

\subsection{Bright star and masking} \label{sec:disc:bs}
Our main experiment aimed to examine how the accuracy of dark-sky flats depending on the observation method. The use of mock objects was merely an auxiliary for implementing the masking process, so we considered only the average number density of celestial objects. 

\begin{figure*}[t!]
\includegraphics[width=\linewidth]{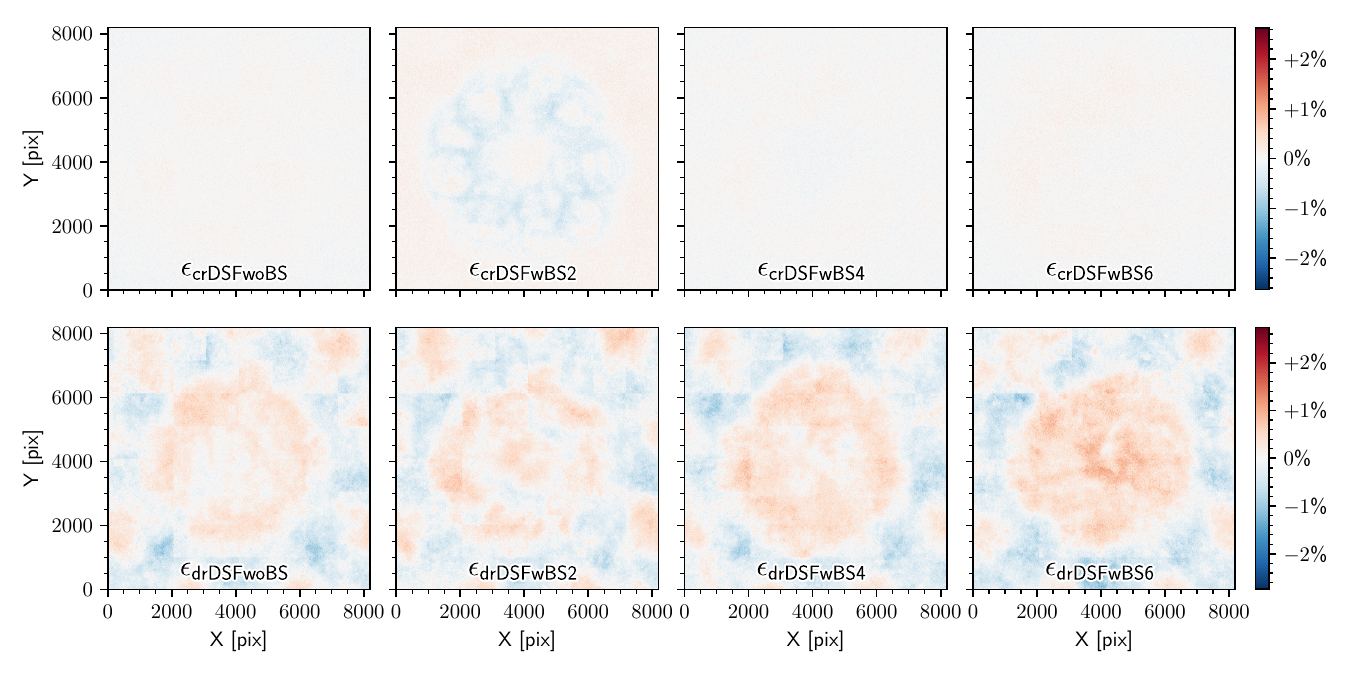}
\caption{Residual maps corresponding to the presence of a bright star and its masking process. The leftmost panel represents the case with no bright star, while the subsequent three panels present the results when the bright star is masked using a circular mask with diameters two, four, and six times larger than the initial segment size. The upper and lower panels show the results for the clear and dirty skies, respectively. The color scales are limited to the $\pm$2\% range of the worst result for each sky. Detailed descriptions of each setup are provided in Table \ref{tab:t2}. \label{fig:f12}}
\end{figure*}

If there are bright stars within the field to be observed, the image may be affected by extended PSF wings and ghosts caused by reflections between the sensor, dewar window, and filter \cite[see][]{2016ApJ...823..123T}. These effects may cause an excess of the sky background level, potentially leading to inaccurate dark-sky flats if the masking process is inadequate. To assess this impact, we simulated a scenario in which a very bright star is present in the observed field. 

A mock bright star was generated using a combination of two Moffat 2D kernels.\footnote{Its shape was derived from an updated PSF profile based on a recent analysis of K-DRIFT P1, conducted after the report by \cite{2022PASP..134h4101B}.} Figure \ref{fig:f11} shows the radial 1D profile of the mock bright star, which has a total brightness of 2.5 mag. The kernel size was set to $\mathrm{8k}\times\mathrm{8k}$ to fully capture the extended PSF wings. In practice, however, the region beyond $R\sim1000$ pixels is unlikely to significantly impact the sky background level due to background noise. To account for reflections from an imaginary dewar window and filter, we introduced four ghost features using Tophat 2D kernels with radii of 60, 120, 180, and 900 pixels. Their brightnesses were set to $1\times10^{-7}$, $5\times10^{-8}$, $2\times10^{-8}$, and $5\times10^{-9}$ times the central brightness, respectively. 

When masking before generating the dark-sky flats, we employed three growing factors to determine the size of the manual circular masks: $2\times$, $4\times$, and $6\times$ the diameter of the initial segment size of the bright star. This resulted in circular mask radii of approximately 650, 1300, and 1950 pixels, as indicated by the green arrows in Figure \ref{fig:f11}. For other objects, fainter than this star but still bright, the mask growing factor of two was applied by default. 

It is worth noting that the outermost ghost feature plays an important role in this experiment since the inner three features would be masked in any case. Since its brightness level is comparable to the sky background level, we can assess the impact of insufficient masking, if any. On the other hand, due to its large size, we can also assess the impact of excessive masking. Specifically, excessive masking has a risk of leading to blank pixels in dark-sky flats, highlighting the need for an observational design that balances the ghost size and the $xy$-shift size. Indeed, for this reason, we have placed the star about one degree away from the reference coordinate to prevent this issue (see Figure \ref{fig:fA-1} in Appendix \ref{sec:app:obj}). 

\begin{figure*}[t!]
\includegraphics[width=\linewidth]{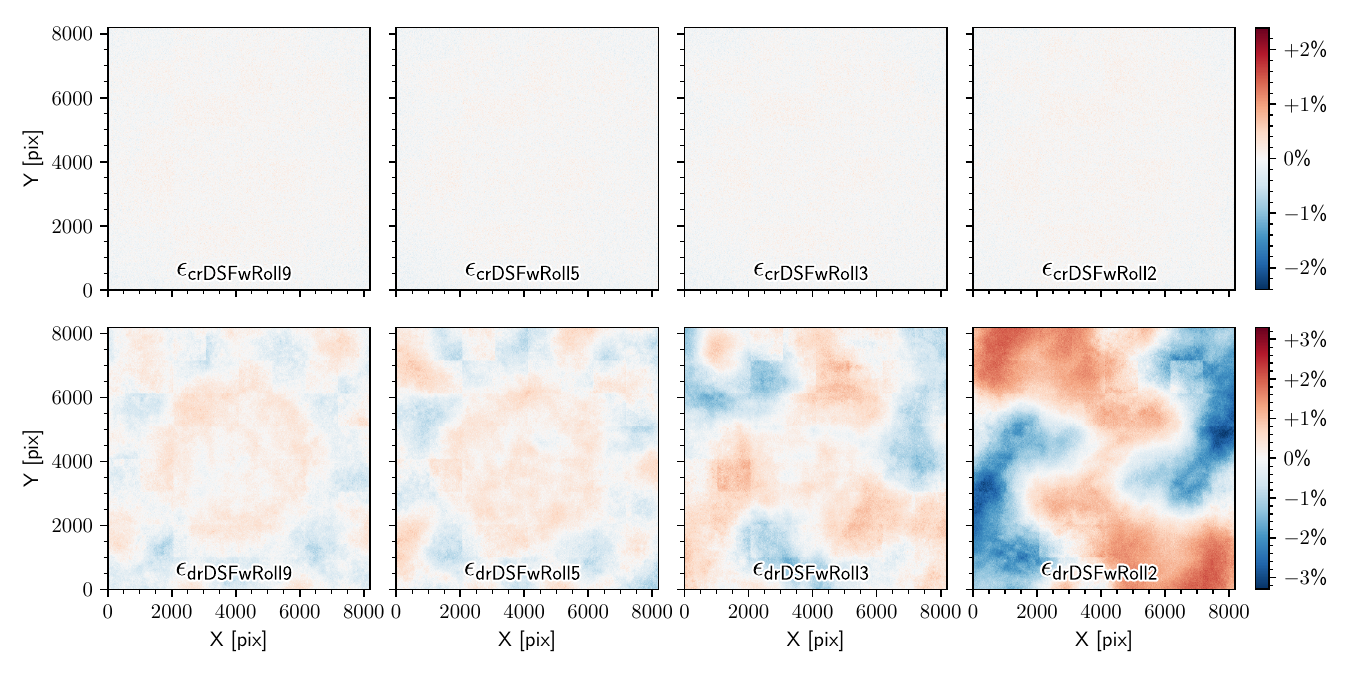}
\caption{Residual maps corresponding to different rolling intervals. From left to right, the maps represent results when the camera rotates by 40$\arcdeg$, 72$\arcdeg$, 120$\arcdeg$, and 180$\arcdeg$. The upper and lower panels show the results for the clear and dirty skies, respectively. The color scales are limited to the $\pm$2$\sigma$ range of the worst result for each sky. Detailed descriptions of each setup are provided in Table \ref{tab:t2}. \label{fig:f13}}
\end{figure*}

Figure \ref{fig:f12} presents the residual maps for four different scenarios involving a bright star. The leftmost panel corresponds to the case without a bright star, while the remaining panels show the results with a bright star present. The most noticeable result appears under clear-sky conditions when the mask growing factor is two. In this case, the ghost feature is insufficiently masked and replicates as the camera rotates, creating a chain-like pattern. Consequently, the spatial peak-to-peak deviation reaches approximately $\pm$0.5\%, compared to only $\pm$0.1\% in other scenarios. 

Interpreting the results under dirty-sky conditions is more complicated. Unlike in clear-sky conditions, a mask growing factor of two does not produce the worst outcome. In fact, when considering only the spatial peak-to-peak deviation, it is 20\% lower than in the case without a bright star. On the other hand, as the mask growing factor increases, the results worsen; a large circular excess of light becomes prominent. This is likely caused by significant DGL contamination during the filling of masked pixels. These results, however, are highly dependent on the shape/brightness of DGL and whether the masking of bright objects is either insufficient or excessive, making it challenging to generalize the findings. 

In conclusion, more careful masking is required to obtain accurate dark-sky flats when observing the sky with very bright stars. If the mask is either too small or too large, the rolling dithering method can have a rather negative impact. To mitigate this issue, it may be necessary to subtract ghost features or deconvolve the PSF. Fortunately, however, there are not many bright stars---especially outside the Galactic plane---that may cause such a significant impact. Thus, the rolling dithering method can still be valuable for observing most sky regions. Investigating optimized treatments for bright stars may be a useful direction for future work. 

\subsection{Different rolling intervals} \label{sec:disc:pa}
All previous experiments demonstrate that the rolling dithering method is effective in obtaining accurate dark-sky flats. While those have focused on the quality of individual images, we now shift our focus to the role of the rolling dithering technique itself. We can ask various questions, such as: What adjustments can optimize the rolling dithering method, and how can we achieve even better results? Among various methodological considerations, the most fundamental approach is adjusting the rolling intervals, which refer to the total number of PAs in an observation sequence or the angular increment of the camera's rotation. 

To select the sets of PAs for the experiments, we considered the following criteria: First, the total number of PAs in an observation sequence must be a divisor of the total number of images to ensure an equal number of images in each PA. Second, there must be no duplicate angles between different sets to maximize diversity in the final coverage. Third, the selected PAs must provide uniform data with azimuthally symmetric coverage while eventually returning to their original angle at the end of the sequence. 

Consequently, we opted four PA sets of [0$\arcdeg$, 40$\arcdeg$, 80$\arcdeg$, 120$\arcdeg$, 160$\arcdeg$, 200$\arcdeg$, 240$\arcdeg$, 280$\arcdeg$, 320$\arcdeg$], [0$\arcdeg$, 72$\arcdeg$, 144$\arcdeg$, 216$\arcdeg$, 288$\arcdeg$], [0$\arcdeg$, 120$\arcdeg$, 240$\arcdeg$], and [0$\arcdeg$, 180$\arcdeg$]. The first set is essentially the same as the default set of the main experiment, where the camera rotates by 160$\arcdeg$. Note that the final set would lead to a near-rectangle coverage, in contrast to the others. If it proves valid, this option could help minimize coverage loss due to rolling dithering. 

Figure \ref{fig:f13} presents the residual maps for the four rolling intervals. Under clear-sky conditions, the results show little or no difference between the rolling intervals. This is because ZL, which governs the clear sky, exhibits an almost linear gradient across the frame. The values of two or more pixels in symmetric positions relative to the reference coordinate along the gradient are consistently averaged out, leaving only the detector's sensitivity variation. For this reason, even in the last case where the camera rotates by 180$\arcdeg$, an accurate dark-sky flat can still be achieved. 

In contrast, the results under dirty-sky conditions show notable differences depending on the rolling intervals. The traces of DGL produce distinct point-symmetric patterns, which vary with the rolling intervals. Additionally, as the total number of PAs decreases, the spatial peak-to-peak deviation increases from $\pm$1\% to $\pm$2.5\%. Nonetheless, the spatial peak-to-peak deviation is significantly reduced compared to the case without rolling dithering. 

At this point, one might wonder if denser rolling intervals could improve the accuracy of dark-sky flats. The answer, however, is not straightforward: in fact, the improvement is minimal once the total number of PAs exceeds five. A more significant challenge lies in the fluctuations, which appear as circular patterns in the residual maps. These fluctuations are unlikely to disappear, regardless of how dense the rolling intervals are. Since this issue is inherently associated with local structures in dirty skies, adjusting the rolling intervals has limitations in improving the accuracy of dark-sky flats. 

In conclusion, rolling intervals do not play such a significant role in obtaining accurate dark-sky flats under clear-sky conditions. This flexibility allows for many options in the rolling dithering strategies, including a 180$\arcdeg$ rotation, which helps reduce coverage loss. On the other hand, under dirty-sky conditions, rolling intervals impact the accuracy of dark-sky flats; denser rolling intervals yield relatively better results. However, the overall improvement remains constrained due to the inherent limitations imposed by dirty skies. 

\section{Conclusion} \label{sec:sum}
LSB structures offer valuable insights into the evolutionary history of galaxies and galaxy clusters, serving as tracers of intergalactic mergers and accretions. Moreover, LSB galaxies, as fundamental building blocks of mass assembly, play a key role in verifying the $\Lambda$CDM paradigm. Therefore, understanding the nature of the LSB regime holds significant importance. 

Despite its importance, most of the LSB universe remains unexplored due to observational challenges. These arise not only from limited light-gathering power but also from systematic uncertainties caused by stray light within the telescope and less accurate data reduction. Although K-DRIFT has a relatively small aperture, its off-axis design minimizes stray light and scattering, making it well-suited for LSB observations. To further augment its observational capabilities, we aimed to improve data processing accuracy---particularly in flat-field correction---by obtaining more accurate flat frames. 

Dark-sky flats, generated by combining science images, are a widely used flat frame for LSB observations. However, the night sky with various underlying light sources is not uniformly illuminated during observations, and fluctuations in brightness can compromise the accuracy of dark-sky flats. To mitigate this issue, we explored methods to obtain dark-sky flats that are less affected by sky fluctuations. 

We first constructed a semi-realistic night sky frame by simulating various light sources, including AG, ZL, DGL, and mock celestial objects. To account for different observational scenarios, we employed two contrasting sky conditions: a clear sky with minimal ZL/DGL and a dirty sky with prominent ZL/DGL. Mock observations were then conducted using two observation strategies: offset dithering, a conventional method involving $xy$-shifts, and rolling dithering, which is akin to offset dithering but adds camera rotation.

After conducting four mock observations with two sky conditions and two dithering methods, we generated dark-sky flats for each dataset and evaluated their accuracy. Our results suggest that the rolling dithering method effectively suppresses sky fluctuations, resulting in more accurate dark-sky flats, with its effectiveness maximized under clear-sky conditions. This improvement may facilitate more precise exploration and photometry of faint, diffuse objects. 

We also performed additional analyses to explore potential factors that could still affect the accuracy of dark-sky flats, even with the rolling dithering method. Our findings are as follows: First, the detector's sensitivity variations do not directly impact the dark-sky flats, as they are pixel-invariant. Second, dark-sky flats may contain non-negligible errors if AG fluctuations are substantial and highly persistent. However, even in such cases, the rolling dithering method helps improve accuracy by introducing additional randomness. Third, the impact of very bright stars on dark-sky flats depends heavily on the masking scheme; too-small masks may leave noticeable artifacts, while too-large masks may eliminate the stars but potentially reduce the number of pixels available for median calculation. Lastly, while denser rolling intervals improve the accuracy of dark-sky flats, the benefit might diminish beyond a certain point. Furthermore, under clear-sky conditions, which allow for flexible rolling intervals, a variety of options are available to suit the purpose of the observation. 

The K-DRIFT survey will leverage the rolling dithering method for in-depth exploration of the LSB universe in the southern hemisphere sky. This approach is expected to achieve an exceptionally deep optical photometric depth, enabling the detection of various LSB structures and galaxies, thus providing crucial insights into the evolution of the universe. 

\begin{acknowledgments}
We are grateful to an anonymous referee for constructive comments and suggestions.
This research was supported by the Korea Astronomy and Space Science Institute under the R\&D program (Project No. 2025-1-831-00) supervised by the Ministry of Science and ICT.

\end{acknowledgments}

%

\vspace{5mm}
\facilities{IRAS, COBE, Sloan}


\software{ZodiPy \cite[][]{2024JOSS....9.6648S}, Astropy \cite[][]{2013A&A...558A..33A,2018AJ....156..123A,2022ApJ...935..167A}, Photutils \cite[][]{2021zndo...4624996B}, Scipy \cite[][]{2020NatMe..17..261V}, Source Extractor \cite[][]{1996A&AS..117..393B}, Astroalign \cite[][]{2020A&C....3200384B}, SWarp \cite[][]{2002ASPC..281..228B}
}



\appendix
\section{2D map of Mock Objects} \label{sec:app:obj}
Employing mock objects is not a critical part of this experimental study. Their primary role is to allocate a plausible proportion of masked pixels, which are involved in the error budget of the median calculation when generating dark-sky flats. Although this masking might introduce small fluctuations in dark-sky flats, our analysis primarily focused on large-scale fluctuations comparable to the frame size. 

\begin{figure*}[t!]
\includegraphics[width=\linewidth]{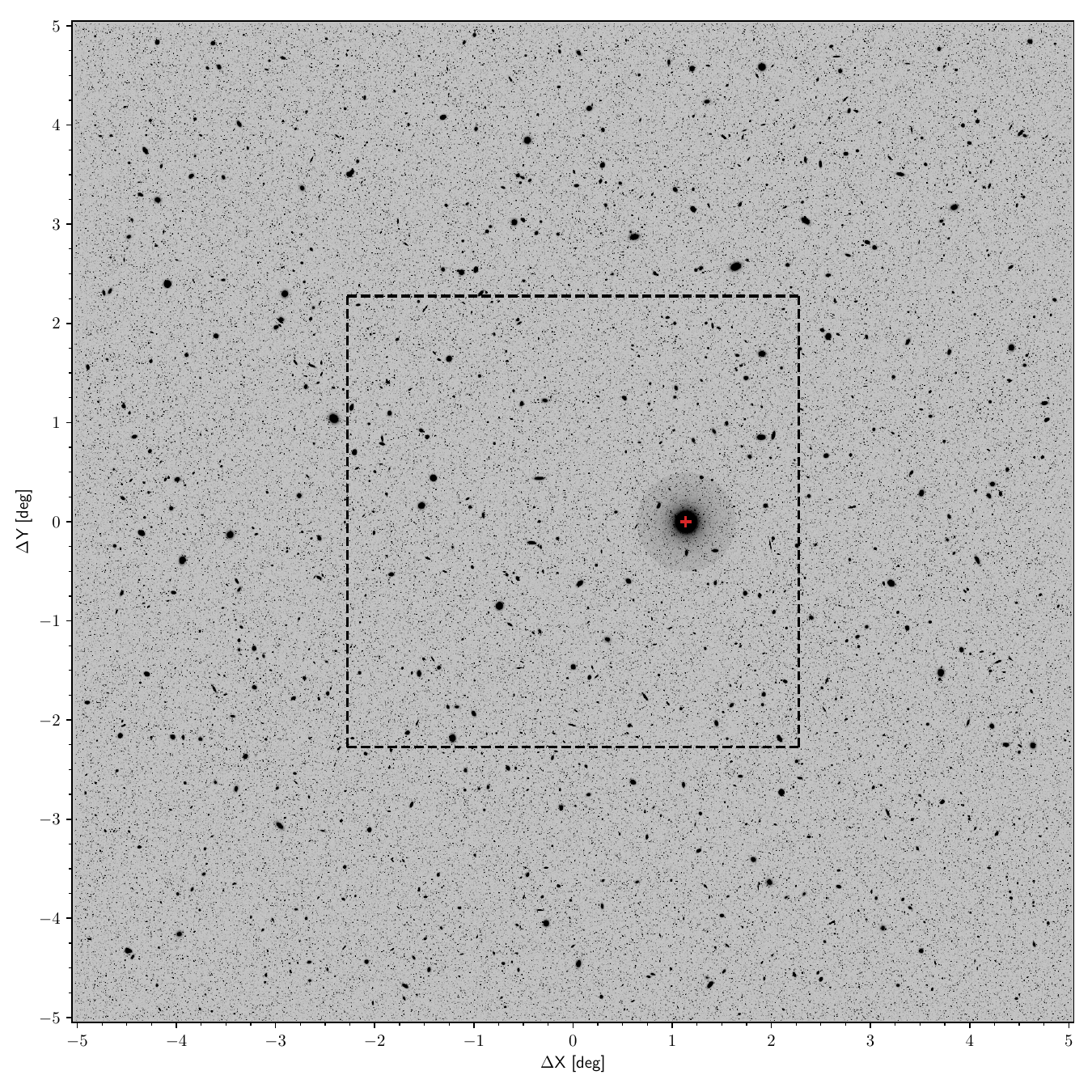}
\caption{Mock object image containing modeled stars and galaxies. A very bright star, marked with a red cross, is included but was only used in Section \ref{sec:disc:bs}. The box with a dashed black line indicates the FoV of a single frame centered at the reference coordinate. For enhanced visibility, random Gaussian noise was added. \label{fig:fA-1}}
\end{figure*}

In fact, superimposing an artificial segmentation map directly onto sky background images without creating mock objects can be an alternative that implements a masking process. However, this approach does not account for variations in local noise or the impact of unmasked light. While a more practical approach is to use images from existing surveys instead of simulating the night sky, this limits the flexibility to modify the characteristics of mock images corresponding to diverse scenarios. Moreover, obtaining uniform, large images as wide as 100 deg$^2$ is challenging. 

Given these issues, we opted to generate a plain frame populated with analytically modeled mock objects. Figure \ref{fig:fA-1} shows the entire image containing the mock objects. Most of their physical properties---except for the number densities and apparent magnitude distributions---are heavily dependent on randomness. This approach is intended to provide a controlled environment for the experimental setup rather than aiming to replicate the real universe precisely. Therefore, a quantitative comparison with the real universe would be inappropriate. 

\begin{figure*}[t!]
\includegraphics[width=\linewidth]{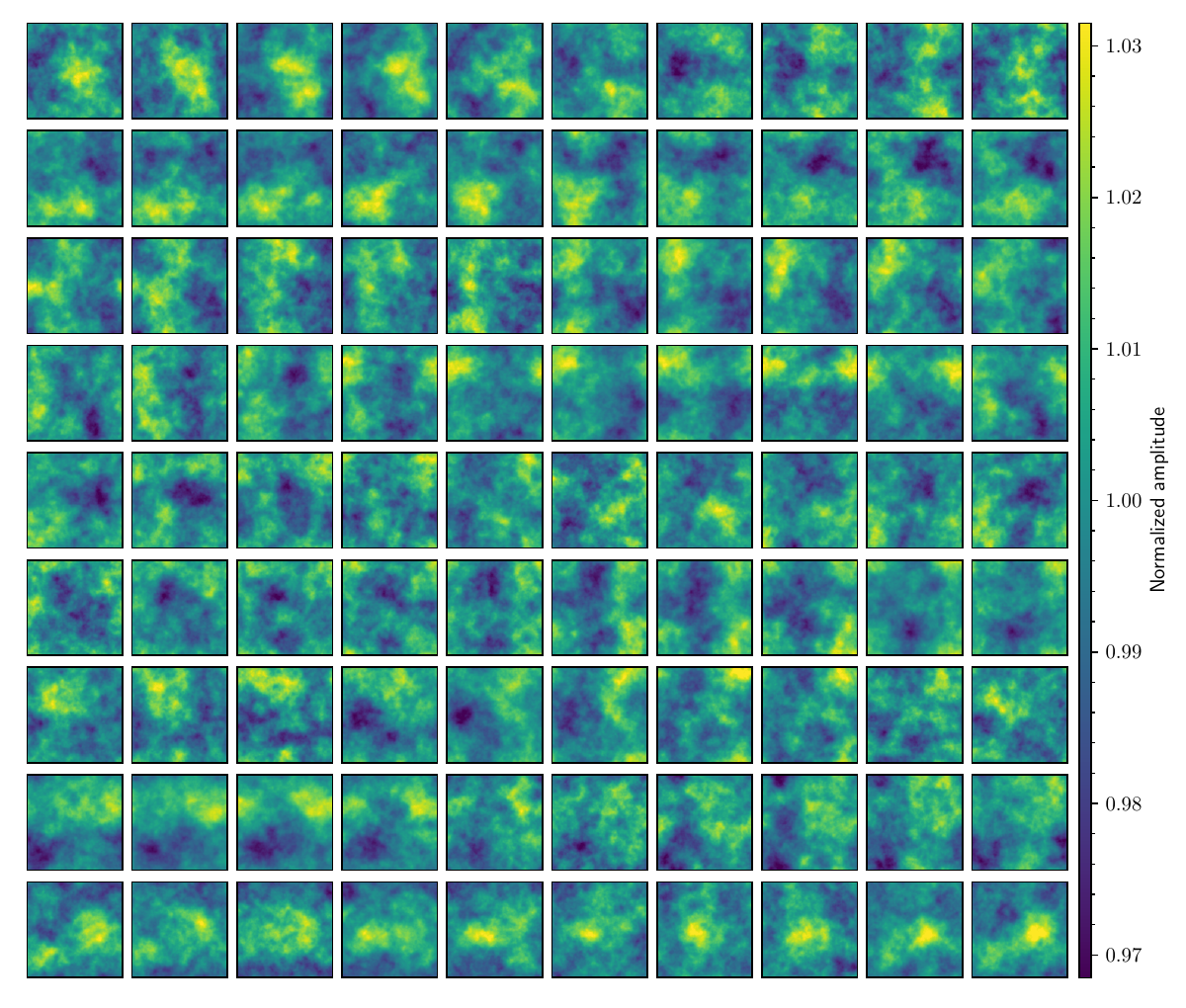}
\caption{Snapshots of AG models created using Brownian noise, with a peak-to-peak deviation of $\pm$3\%. These images imitate the temporal fluctuation of AG over an hour, displayed from top left to bottom right. Each row presents AG models for acquiring mock images with ten exposures at each PA. \label{fig:fA-2}}
\end{figure*}

\section{Models of fluctuating AG} \label{sec:app:brown}
Simulating realistic AG fluctuations is challenging, as they involve multi-wavelength emissions across different atmospheric layers. Their complex timescale variations further complicate the process. One goal of this study is to examine the impact of AG fluctuations on dark-sky flats through mock observations. Thus, we adopted Brownian noise as a practical and efficient approach for modeling fluctuations in the brightness of the sky background between frames. 

Pink noise, also known as fractal noise, is often used to simulate noise in natural processes. However, its dominance of high-frequency variations results in excessively fine structures. In contrast, Brownian noise, which exhibits higher intensity at lower frequencies, produces more clumpy structures. Moreover, it is a random walk process where each step depends on the previous one, making it more suitable for mimicking continuously fluctuating AG. 

Figure \ref{fig:fA-2} shows 90 snapshots of AG models used to generate mock images affected by AG fluctuations. Noise patterns in the same row exhibit gradual variations, maintaining continuity. On the other hand, patterns at the end of one row and the beginning of the next do not have similar patterns. In principle, a continuous noise map should be generated first and then rotated to match the camera's rotation every ten exposures. However, since rotation inherently alters the pattern, we instead generated ten independent noise maps using a new random seed as a practical alternative. 


\bibliography{ms}{}
\bibliographystyle{aasjournal}



\end{document}